# Local Meaning Structures:

## A Mixed-Method Socio-Semantic Network Analysis

Nikita Basov
*Centre for German and European Studies*
*St. Petersburg State University*

Wouter De Nooy
*Department of Communication Science*
*University of Amsterdam*

Aleksandra Nenko
*Institute for Design and Urban Studies*
*ITMO University*

**Abstract**

This paper proposes a mixed-method socio-semantic network analysis of meaning structures in practice. While social and institutional fields impose meaning structures, to achieve practical goals field participants gather in groups and locally produce idiocultures of their own. Such idiocultures are difficult to capture structurally, hence, the impact of practice on meaning structures is underrated. To account for this impact, we automatically map *local meaning structures*—ensembles of semantic associations embedded in specific social groups—to identify the focal elements of these meaning structures, and qualitatively examine contextual usage of such elements. Employing a combination of ethnographic and social network data on two St. Petersburg art collectives, we find the seemingly field-imposed meaning structures to be instantiated differently, depending on group practice. Moreover, we find meaning structures to emerge from group practice and even change the field-wide meaning structures.

# Introduction

In this paper, we accentuate the relevance of collective practice in small groups to meaning structures within a social field. Structural approaches in institutional and field perspectives conceptualize meaning structures as a duality of meanings and social structure (Mohr, 1998) and hold that meaning structures are imposed, for instance, by 'objective relations' (Bourdieu & Johnson, 1993; Bourdieu, 1996a) or by institutions (Mohr, 1994; 2000; Mohr & Neely, 2009) in such a way that the objective relations between positions in social fields correspond more or less to relations between cultural categories. However, there are different fields (Bourdieu & Johnson, 1993; Bourdieu, 1996a) and different institutional orders in society (Friedland & Alford, 1991). Individuals are able to play on gaps and inconsistencies between imposed cultural constructs (Bourdieu, 1990; Friedland & Alford, 1991; Bourdieu & Wacquant, 1992). Joining in groups, supporting each other, interacting, and engaging in shared practice, agents create group-specific cultures, or 'idiocultures' (Fine, 1979) (see also Yeung, 2005; Fine, 2012). Implementing imposed meaning structures associated with different fields or positions, groups often adjust them to their local contexts (De Nooy, 2003). As a result, meaning structures are not merely reproduced, but are instantiated locally or even emerge—as groups strive to self-identify and to take certain positions in the field.[1] A structural account of culture not considering local meaning structures is incomplete, limiting its capacity to explain the evolution of meaning structures in fields and the interaction between cultural structures of different fields and positions.

However, a methodological gap is involved. In organizational and institutional settings, typical for the seminal structural studies of culture, directives and other formal statements explicitly impose meaning structures (e.g., Mohr, 1994). Here, official documents can be coded and then analyzed to investigate imposed meaning structures (Mohr, 1994; 1998; Mohr & Lee, 2000; Mohr & Neely, 2009). This, however, is not the case if we want to capture emergent meaning structures. Neither, we argue, apply the approaches imposing categorizations by researchers or 'ground truth experts', who are field participants themselves, would that be coding of textual data (McLean, 1998; 2007) or using survey data, where informants choose from a set of categories pre-defined by the researchers (Yeung, 2005). Cultural practice often seems irrelevant or trivial to an outsider who attempts categorizing it logically (Bourdieu, 1990, p. 90), hence the limitations in analyzing the encoded cultural practice data statistically (McLean, 2007, pp. 114-119). To account for the emergent meaning structures, the empirical approach must be adjusted.

---

[1] While this study focuses on meaning structures in the framework of social fields, we keep in mind that group culture is more than classifications related to fields. It is also shared stories, common points of concern, relational expectations (White, 1992; Fuhse, 2011; Godart and White, 2011), and so on. These are not only molded by the interplay of fields and practice, but also affected by socioeconomic and neighborhood differences, personalities, and many other factors.

We propose a mixed-method socio-semantic approach based on an inductive culture mapping technique (Carley, 1994; Lee & Martin, 2015), i.e., we start by capturing all collocations of words in verbal expressions of individuals. Simultaneously, we remain within the framework of the duality of social structure and culture (Breiger & Puetz, 2015; Lee & Martin, 2018): Social structure is measured as cohesive clusters in networks of interpersonal interaction and we view semantic structures embedded in *specific* social-structural clusters as meaning structures. Our core assumption is that meaning is constituted by the social context, so associations between words that are shared by members of the same social group carry the traces of this group's practice and idioculture. Using a technique similar to Lee & Martin (2015) who compare semantic associations of individual philosophers of the Frankfurt school, we proceed by selecting words shared within a particular small group using verbal expressions produced by its members. Combining cultural and social structures this way, mapping meaning structures boils down to comparing word associations among groups in a social network.[2] We quantitatively find the focal points in these maps—to start our analysis from the center of the shared meaning structures. To understand (*Verstehen*) these focal meaning structures in the practical context of the group, we use ethnographic data and qualitatively examine how the corresponding associations are used in practice. Simple and straightforward as this method may be, such an approach has, to the best of our knowledge, not yet been proposed (compare, e.g., with Erickson, 1988; Schultz & Breiger, 2010; Vaisey & Lizardo, 2010; Nerghes et al, 2015; Basov & Brennecke, 2017; Martin & Lee, 2018; Godart & Galunic, 2019).

To substantiate our theoretical argument and test our method, we analyze meaning structures and interpersonal relations in artistic collectives: groups of creatives engaging in collective production and representation of artworks. Artistic collectives are a suitable case in point. On the one hand, from the work of Bourdieu (Bourdieu & Johnson, 1993; Bourdieu, 1996a), quite a lot is known about the forces operating in fields of contemporary art and they are known to impose meaning structures on their participants. On the other hand, in artistic fields symbolic capital is strongly associated with originality and new meanings often emerge in artistic circles (Farrell, 1982; 2003). In particular, we study two collectives of artists in St. Petersburg, Russia, each uniting members who cooperated intensely for a long time and developed recognizable practices. This empirical setting allows both for similarities of meaning structures across the collectives associated with the common field and for differences between them attributable to the effect of collective-specific practices. We examine an extensive dataset including a corpus of texts, social network data, and ethnographic data gathered during the two years of our ethnographic studies of the collectives.

---

[2] For a different approach to comparing meaning structures of different groups see Yeung (2005) and for a comparison of cultural meanings across social positions see McLean (2007, pp. 114-119).

## Meanings imposed by fields and meanings emerging in group practice

In their seminal work, Friedland & Alford (1991) proposed a duality perspective on society as an inter-institutional system that both sets material patterns of activity and develops symbolic systems categorizing individuals and activities, thus ordering reality and providing sense to things. According to this perspective, field-specific institutional logics impose both social structure and the meaning of positions within the structure (Breiger, 2000; Mohr, 2000; Breiger, 2005; Thornton et al, 2012). It has been shown empirically how institutions, for example, welfare organizations (Mohr, 1994), universities (Mohr & Lee, 2000) or city administrations (Meyer et al, 2012), group individuals both materially and symbolically, bringing them together, ascribing similar identities to them, and imbuing them with certain meanings. This leads to a duality of social positions and meanings: Distinct meanings are institutionally attached to individuals or groups occupying particular social positions.

However, there are multiple institutional logics in society and the degree of consensus on classificatory principles, according to which meanings are assigned to positions, may vary (Mohr, 1994; Bian et al, 2005), which results in inconsistencies between meanings and social positions. As Friedland & Alford (1991) argue, such gaps and overlaps are often used by individuals and groups to manipulate and reinterpret meanings imposed by institutions, playing with different material, historical, or social contexts and choosing institutional logics, role identities, and activities better suiting their current needs (compare with the notion of switching between netdoms in White (1992); Godart & White (2010)). For example, someone promoting a subordinate who is also one's son is acting as a parent who is expected to support the child instead of acting as a manager who is expected to treat all employees equally. Throughout this process individuals attempt to understand themselves and the world. There is a "meaningfulness of relational systems that enable and constrain action" (Mohr & Rawlings, 2010, p. 16).

Bourdieu's field theory has a similar dual approach to the relation between social structure and meaning. Bourdieu theorizes fields as "spaces of objective relations that are the site of a logic and a necessity" (Bourdieu & Wacquant, 1992, p. 97). Fields correspond to areas of social life, like art, academia, and religion, which consist of objective relations as well as shared understandings of what goes on in the field (Bourdieu, 1983; Bourdieu & Johnson, 1993; Bourdieu, 1996a). Objective relations fundamentally structure fields according to both social properties held by agents in certain positions and categories of perceptions used by the agents (high/low, masculine/feminine, large/small, etc.). The latter constitute cultural categorizations shared by members of the field, especially by those who occupy similar positions (Bourdieu & Wacquant, 1992). This way, objective relations impose meanings on members of a field.

Neither institutional theory nor field theory describes the relevance of interpersonal interactions to the meaning structures shared by members of an institution or field. Meanwhile, there are good reasons to believe that interactions matter. De Nooy (2003, p. 323) argued that interpersonal relations such as interaction are relevant to field theory because, on the one hand, interpersonal interactions "mediate and transform the effect of objective relations". On the other hand, individuals in one field may engage in interactions with individuals in other fields, "bringing to bear properties and qualifications characteristic of another field", giving rise to new meanings that comprise hybrid classifications, stigmata and identities, in turn affecting the structure of cultural categories. "In the process new symbolic distinctions and values are being created or existing ones are being reaffirmed or discarded".

The relation between interaction and symbolic production proposed by De Nooy is just one among a number of perspectives that link networks of interpersonal relations to culture. As DiMaggio stresses, "network analysis is the natural methodological framework for empirically developing insights from leading theoretical approaches to cultural analysis" (DiMaggio, 2011, p. 286). Attitudes, opinions, and tastes were shown to reproduce through interactions and therefore to rely on interpersonal ties (Carley, 1986; McLean, 1998; 2007; Godart & Galunic, 2019). More recently, similarity in perceptions and meanings has been argued to affect interpersonal relations (Vaisey & Lizardo, 2010; Dahlander & McFarland, 2013). Overall, culture and interpersonal ties are increasingly seen as mutually constitutive and therefore, dual (Breiger, 2000; Mohr, 2000; Basov & Brennecke, 2017; Martin & Lee, 2018; Godart & Galunic, 2019).

This literature suggests an intimate relation between interaction and meaning. From the perspective of meaning structure as a combination of social structure and meaning, we may expect individuals who are structurally grouped by interpersonal interactions to strive towards shared meanings. In particular, shared meanings are to be expected in small groups, known to produce idiocultures of their own (Fine, 1979; 2012). In their verbal expressions and discussions, they express local meaning structures guided by the characteristics of their joint practice. As a consequence, the sense of cultural elements varies across meaning structures of different social groups within the same field. We expect local meaning structures to be especially vivid where different fields or institutions collide, for example, when members of different fields interact in heterogeneous groups.

Importance of group interaction to meanings has been emphasized by a rich research tradition of studies of art including Becker (1982) that has shown how repetitive and intense interactions among artists develop into conventions and correlated schemes of behavior in an art world. According to the collaborative circles theory (Farrell, 2003), engagement in discussions and mutual support in their coordinated practice enable groups of artists to develop and promote unconventional shared 'visions', which are the core of an art group's culture. These visions guide the collective work of artists, coordinate their artistic styles, their techniques and topics. Supporting each other emotionally throughout joint communication and practice, group members are able to probe the boundaries and gain new insights, reinforce them and promote them more broadly (Farrell, 2003), that may later become part of institutionalized cultural constructs (see Godart & White, 2010).

## An inductive mixed-method socio-semantic approach to meaning structures

Highly regulated meaning structures, imposed by institutions, have been fruitfully investigated by coding words assigned to individuals in similar social positions and words corresponding to institutional practices into a limited set of categories; then, the relations between the two types of categories can be formally analyzed (Mohr, 1994; 1998; Mohr & Lee, 2000; Mohr & Neely, 2009). Small group cultures are far less regulated. There, one should not expect expressions to be "succinct and formalistic, <...> loosely coupled from the demands of conforming to objective practice" (Mohr & Neely, 2009, p. 220) and to converge on a limited set of conventional categories that a researcher or an expert can detect from the outside of the group. Any categorization would imply that a researcher imposes certain categories on the textual data (for an extensive discussion see Lee & Martin, 2015), which is problematic if we want to investigate how individuals shape meaning structures while interacting in practice. As shown in the test of Bourdieu's theory conducted by McLean (2007, pp. 114-119), who encoded thousands of Renaissance patronage letters and statistically related them to the structure of social positions and the social network structure, cultural practice cannot be fit into a small number of keywords without losing substantial information. Consistently with the arguments of the new institutional and field theorists discussed above (Bourdieu, 1990; Friedland & Alford, 1991; Bourdieu & Wacquant, 1992), McLean (2007, pp. 118-119) concludes that one of the main reasons his analysis does not fully support the Bourdieuan argument is that "agents skillfully manipulate their social structural profile, foregrounding certain social positions and backgrounding others". When dealing with texts produced in everyday interaction, coding inevitably distorts their cultural meaning in order to fit them into researcher's classification system: "It is one thing to predict how many particular works of art or music a person would recognize on a survey based on her social structural position (Bourdieu, 1984). It is another thing to classify a more complex piece of cultural production like a letter and to identify higher-order cultural components in it" (McLean, 2007, p. 117). This is perfectly in line with the argument by Bourdieu (1990, p. 86) that "practice has a logic, which is not that of the logician". McLean concludes that encoding of textual data that comes from cultural practice biases subsequent statistical results based on these data. One can hardly imagine a coder more knowledgeable of her or his subject than McLean is of the Renaissance patronage letters he studied for decades. Hence, the problem is unlikely coder selection. The coding approach itself is not a proper technique to formalize culture emerging in practice.

McLean is also right in pointing out that closed-ended survey questions are no less problematic than coding when trying to capture the complexity of cultural practice. Even more, such questions exclude this complexity from the dataset per se—by imposing culture from the very beginning, i.e., when a question is formulated, hence filtering the input. When the data were ordered at the input stage, perhaps, the models converge more easily and the illustrations of the meaning structures are easier to comprehend without prior ethnographic knowledge of the field, but it does not mean the local culture is well-captured. Rather, it is well-hidden: The 'messy' part of culture emerging in practice simply does not make it into the dataset. Unlike with data from closed-ended survey questions, statistically analyzing the encoded data one at least has a chance to discover that something is being missed—like McLean did.

Following the lead of Lee & Martin (2015) and the recent movement in the formal analysis of culture (see also Mohr, 1998; Mohr & Bogdanov, 2013; Bail, 2014), to embrace the complexity of culture in practice we seek an inductive strategy swapping the qualitative and the quantitative stages of analysis. That is, our starting point is to map all cultural constructs represented in the texts (e.g., textual artworks, narrative interviews, and group conversations) informants produced as freely as possible, instead of starting by selecting what is important *to us*.

How can we map culture from verbal expressions in the practical context of a particular social group? The tradition set up by Geertz (1973) and social constructivists (Berger & Luckmann, 1966) suggests viewing culture as a repertoire of elements shared among members of a social group that characterize this group. With respect to verbal expressions, this implies meaning in the linguistic-philosophical (Wittgenstein, 1953) and socio-linguistic (Labov, 1972) sense: The meaning of words relies on a particular social context in which they are used. Hence, we are evidently less interested in the dictionary meaning and take semantics in the strict linguistic sense for granted.

To go beyond dictionary meaning, we look at textual structures larger than a single word. We focus on how individuals associate (collocate) words in their verbal statements, e.g., within a sentence (Sinclair, 1991). This approach aligns with co-occurrence-based semantic network analysis (Carley, 1994; Diesner, 2013; Lee & Martin, 2015; Nerghes et al, 2015), a technique starting with a corpus of texts rather than with a set of predefined categories and producing maps of associations between word stems—traditionally named 'concepts'—as they appear next to each other within a certain interval in these texts. Because such co-occurrences are rarely controlled by the speakers, this technique is capable of capturing the associations put forward unconsciously, among others.

Applying the collocation technique to texts by individual artists separately, we get concept associations specific to different artists in particular contexts. For example, all members of artistic collectives use the concept *art*. However, their understanding of art may differ, which can be revealed by the relations between this concept and other concepts in their discourses. Some may associate *art* with *transformation*, *culture*, or *history*, while others associate it with *canvas*, *brush*, or *palette*. This signals that *art* is something different to them. Thus, we capture the polysemy that is inherent to a logic of practice that "never explicitly limits itself to any one aspect of the terms it links, <…> exploiting the fact that two 'realities' are never entirely alike in all respects, but are always alike in some respect, at least indirectly (that is, through the mediation of some common term) <…> like the keynote to the other sounds in a chord, to the other aspects, which persist as undertones" (Bourdieu, 1990, p. 88).

More conducive to the perspectives of individuals who express themselves in practical contexts, word co-occurrence-based culture mapping has at least one important drawback: Concepts may co-occur for lots of reasons, ranging from grammatical rules to slips of the tongue, hence the slightly derogatory 'bag-of-words' label for co-occurrence-based approaches. In contrast to the analyses that link culture to social structure after choosing 'keywords' manually (whether during research design (Yeung, 2005) or during analysis (McLean, 1998)), our approach requires locating the important words in the inductively produced 'hairball' maps consisting of tens of thousands of concepts and associations between them. We do that by building the account for social structure into the cultural data formalization procedure: Information on who produced which text and the social relations among them is used to select the concepts and their co-occurrences. This turns our approach into socio-semantic network analysis (Roth & Cointet, 2010; Basov et al, 2017; Saint-Charles & Mongeau, 2018).[3]

We computationally select associations that different individuals share. If the individuals are structurally similar owing to the pattern of their interactions or membership of the same collective or field, the concept associations they share indicate the meaning structure of the collective or field. This way, we relate meanings to the social structure of interpersonal relations and field positions while at the same time substantially reduce complexity of the concept maps.

---

[3] For alternative socio-semantic approaches, relating social networks to semantic networks *after* the latter were constructed, see Basov and Brennecke (2017) for multiplex networks, and Lee and Martin (2018) and Godart and Galunic (2018) for multilayer networks.

Then, the formally detected intersections in concept associations can be inspected at the field level, presumably representing meaning structures imposed by the field. We can also examine meaning structures specific to artistic collectives and subgroups within these collectives.

The proposed aggregation approach is in line with Bourdieu's practice theory: "the cumulation and juxtaposition of relations of opposition and equivalence which are not and cannot be mastered by any one informant, never in any case at the same time, and which can only be produced by reference to different situations, that is, in different universes of discourse and with different functions, is what provides the analyst with the privilege of totalization" (Bourdieu, 1990, p. 82). Hence, meanings are only partially expressed in single practical situations and practice is to be captured across situations and individuals.

Unlike another inductive aggregation approach—topic models (Blei et al, 2003; DiMaggio et al, 2013; Mohr et al, 2013)—that puts hundreds of separate concepts into an arbitrarily selected number of piles[4], co-occurrence-based concept associations mapping traces systematic joint usage of concepts and yields networks of relations between specific concepts based on that. These structures of associations can be compared. For instance, if in one group the concept *art* co-occurs with concepts such as *transformation*, *culture*, and *history* and in another group—with concepts such as *canvas*, *brush*, and *palette*, we can say the first group views art as a means of social change, while the second one—rather as creation of an art piece. Moreover, because the number of co-occurrences is counted, relative strength of concept associations can be compared. For instance, if *art* co-occurs 100 times with *politics* and only 10 times with *exhibition*, we can say that a group rather views art as political means than as production of aesthetic objects for galleries. Furthermore, we can find specific concepts that are most strongly associated with other concepts and locate the ones having many unique associations.[5] The more polysemic a concept is, the more it is elaborated and the greater the chances for this concept to connect across different field positions, making it central for a greater number of participants (see Yeung, 2005, p. 398). The strongest associations and the concepts with many associations can be compared across groups, to distinguish between their focal meaning structures. This opportunity allows us to go beyond topic modeling, which does not tell which cultural elements are more important to those expressing them and how particular elements are related to other elements.

---

[4] Topic models require researcher to impose the number of topics for the algorithm to find; this number is difficult to justify (see, e.g., Bail, 2014). Similarly to coding (including coding as part of supervised topic modeling (Blei and McAuliffe, 2010)), this does not suit the purpose of capturing meaning structures in practice.

[5] Displaying the concept associations as a network, it is tempting to read sequences of associations as sentences or stories, similarly to the chains of implication for personality characteristics and relationship qualities in Yeung's analysis of communities' meaning structures (Yeung, 2005, pp. 405-407; see also Bearman & Stovel, 2000). In all honesty, we initially yielded to this temptation. But when we took into account the textual context of the associations, we soon discovered that it is misleading to interpret indirect associations in this type of semantic networks. As a consequence, we doubt that measures

The central concepts and frequent associations shared in a certain group signal where to look for the focal meaning structures of that group. It can also help us in finding the meaning structures that distinguish a certain group from the rest. However, only qualitative analysis is capable of understanding (*Verstehen*) these meaning structures in their practical context. To interpret group-specific meaning structures we apply and advocate a mixed methods approach. Namely, we examine all the quotes utilizing the focal meaning structures of the groups and strive to understand them in the context of collective practice by checking with the other ethnographic materials.[6] Similarly, group structure within a collective derived with social network analysis should be validated against the perceptions of groups and roles by the members of the collective taking into account the context of their practice. This can only be done by a human interpreter well-aware of the collective's practice (i.e., field researcher who collected the ethnographic data).[7]

---

accounting for overall network structure, that is, the core of social network analysis, are useful for analyzing semantic networks of direct collocation when focusing on meaning structures. In this kind of analysis, our network visualizations are merely convenient ways of jointly presenting dyads of concepts that signal meaning structures. As a consequence, we limited ourselves to usage of dyad-based network statistics, such as degree centrality in this paper. Elsewhere, we experiment with statistical models that focus on local micro-patterns of social ties and concept associations (references to authors' own work).

[6] In addition, there are good technical reasons for such a qualitative inspection. When meaning resides in context of a sentence it cannot be teased out automatically. For instance, consider differences in the attitude of the speaker to political performances: 'all performances must be political performances' in contrast to 'political performances are senseless'. Computer algorithm maps an association political—performances out of both. Hence, a proper interpretation of the semantic networks requires manual checking the actual quotes from which concept associations were taken.

[7] This does not dismiss quantitative analysis, which allows tracing subliminal yet stable semantic associations an interpreter can hardly pinpoint in the flow of ethnographic data that comes from a diversity of situations. In our experience of such quantitative analysis, when meaning structures are mapped, both field researchers and group members confirm they make sense. But neither researchers nor members are able to point at these meaning structures in the narratives prior to mapping (see also Bail, 2014; Mcauliffe & Blei, 2008).

## Data

This study investigates two artistic collectives located in St. Petersburg, Russia, encoded as C and Z. Both were founded in 2003, both do contemporary visual art, and are characterized by intense interaction, strong interpersonal ties, and the joint artistic practice of its members. Both of the collectives are highly innovative and have introduced their distinctive practices and themes into the Russian art field. Yet, the two are different in organization, the educational and cultural backgrounds of members, understanding of art and its tasks, spatial embedding in the city space, and artistic styles. Note that Z includes only individuals occupying positions within the Russian contemporary art field, while C combines members from two fields: the art field and the academic field (further, we refer to them as 'Artists' and 'Academics').

Artistic collectives are highly informal. In particular, membership in Z is open and flexible, while C members collaborate closely with other artists and intellectuals. In this analysis, we only focus on the participants with stable membership, which implies long-standing involvement in interaction and joint practice of the collectives, which we assume a prerequisite for them to engage with meaning structures of the collective.

Z includes 11 core members aged 20 to 65, mostly young persons, connected by many years of friendship and collaboration. The distinctive feature of the collective is its exhibiting strategy. It 'occupies the bodies of other cultural institutions' and acts as a 'nomadic gallery' invading unconventional parts of conventional artistic spaces (e.g., a wall of a theatre or a corridor of an art-center where Z artists exhibit every two weeks), presenting works in everyday city spaces (e.g., a grocery shop or a public toilet), and performing in the city streets. Irony towards market-oriented gallery art is, perhaps, the main message communicated by this collective. Hence, the works often parody or mock commercially successful artworks, while being created in a trash manner (Fig. 1). Some senior members were very well known already in the Soviet underground art scene, and some of the junior ones are increasingly successful in Russia and internationally.

_______________________________

Figure 1 about here

_______________________________

C consists of nine core members aged 30 to 50 (the majority of members being closer to the latter age), most of whom are tied by friendships lasting for decades. Thematically, the collective focuses on political issues and draws on ideas of the leftist critical theory (e.g., Gramsci, Brecht, Marx, Althusser). C is acknowledged in Russia as one of the agenda-setting politically engaged artistic collectives and is integrated well into the international art scene, having gained considerable success there among grant-givers, art professionals, and publics. Artwork formats (Fig. 2) are mixed and predominantly incorporeal: textual (e.g. a newspaper named after the group name), performative (e.g. musical shows—the 'Songspiels'), and actionist (e.g., artistic demonstrations). The members are affiliated (often in quite high posts) with a variety of organizations in St. Petersburg and Moscow—educational and research institutions, publishing houses, museums, theatres, centers for contemporary art.

_______________________________

Figure 2 about here

_______________________________

The data were collected in 2011 and 2012. In line with our purposes, we distinguish between two main types of data: textual and social network. To map interpersonal relations between the members we used a sociometric survey providing them with the list of all other core participants and asking: 'How often do you interact?' The respondents were asked to choose from among five options for each other member: almost never; 1 or less/month; 2–4 times/month; 5–14 times/month; 15 or more times/month. Seventeen out of twenty core members responded to the survey resulting in a response rate of 85%. If members evaluated the frequency of their interactions with each other differently, we used the median of their reported interaction frequencies.

Our main method of textual data collection was in-depth interviews with the artists. The interviewers asked up to 50 open-ended questions from a standard list, such as: 'How did your group form?'; 'How would you describe your relations with other members?'; 'What do you do together with other members?'; 'What is art to you?'; 'How would you describe your artistic style?'; 'Do you sell your works and how?'; 'What does viewer's reactions on your works mean to you?'; 'How do you communicate with the audience?'; 'How would you describe the artistic scene?'. The 17 interviews lasted from 55 minutes to 5.5 hours, depending on the interviewee's endurance. We also collected texts written by members of the collectives having clearly identifiable authorship. The written texts include webpages, blogs, newspaper articles, poetry, novels, and so on. Finally, we conducted 14 visual ethnographies in artistic studios and at exhibitions lasting for about two hours and registering the processes of individual and collective artistic work, informal communication between members, encounters of the artists with other creative professionals and broader publics, and preparation of exhibitions. Longer expressions in conversations witnessed during these observations were transcribed to produce additional textual data. For each of the members we managed to collect at least one type of textual data and at least two types for most of the members.

We started by using all texts collected for a core member—the member's corpus—to map this person's semantic network. The following procedure was applied to each corpus separately. Firstly, stop words deemed irrelevant to meaning structures such as articles, particles, prepositions, pronouns, rhetorical, intermediary, and bridging words were marked. Secondly, we applied Porter's stemming procedure for the Russian language (all textual data was collected in Russian) to aggregate variants of the same word, e.g., singular and plural forms, to their stems by removing suffixes (Porter, 1980).[8] These generalized forms of aggregated words are the nodes in our semantic network, referred to as 'concepts' (see Carley, 1986; 1994; Diesner, 2013; Lee & Martin, 2015; Nerghes et al, 2015). Table 1 presents the number of concepts per member per text type. These counts show the amount of textual information that we have for each member.

Finally, we created an undirected link between every two concepts if the words from which the concepts derive appeared at least once next to each other in the member's texts, i.e., unless separated by another concept(s) or by a stop word(s). The concepts and links between them constitute the undirected semantic network of a member. For example, if member A's textual corpus consisted only of the phrase 'Dima makes bad performances: Dima's performances don't attract people', the semantic network of A would consist of six concepts, namely *Dima*, *performance*, *bad*, *make*, *attract*, *people*, and five associations *Dima—performance, make—bad, bad—performance, Dima—make, attract—people*.

---

[8] We refrain from using thesauri to collapse synonymous words for the very same reason we refrain from taking words in their dictionary meanings or coding: to avoid imposing meanings. Dictionary synonymy is not a sociolinguistic one whereas thesauri development by researchers is just another way of biasing the formalization procedure with researchers' interpretations.

We insist on such, most restrictive, approach that considers only immediate collocation of concepts in text as a co-occurrence for the sake of reliability. Other studies (see, for example Lee & Martin, 2015) use (1) a greater 'window size', i.e., the maximum amount of other words between the focal words given which the focal words are still considered connected, or connect all words within a paragraph, or even within the whole document, and (2) ignore 'unimportant' stop words between the concepts when calculating the window size. However, many years of trials and errors in semantic mapping of texts in different genres in the Russian language have taught us that in practical settings one can be sure that words are positively associated only if they are put immediately next to each other. The passage in the previous paragraph exemplifies, why. Even using a window size of three, or using a window size of two but ignoring stop words when calculating window size, would yield misleading output: We would connect *performance* and *attract* in the statement 'performances don't attract' (in Russian: 'перформансы не привлекают')*,* associating the words that are, in fact, dissociated. Although using such a restrictive approach we definitely lose quite some additional information, this increases the chances that the associations we do map are existing and positive associations.

The number of associations per semantic network is provided in Table 1. It shows that the size of the corpus of concepts varies among members as well as the source. For some members we have more textual material and larger semantic networks with more associations than for other members. We think, however, that this is as it should be in an inductive approach. Members who are more vocal are more likely to express and perhaps shape the communis opinio of a social group.

_______________________________

Table 1 about here

_______________________________

In line with our approach, the subsequent analysis accounts only for concept associations shared within or across the collectives; Table 1 includes the numbers of associations shared with at least one other member of the two collectives. For example, if semantic networks of members A and B both contain the association *make—performance*, we consider this association as shared between A and B. Shared associations are quite rare in comparison to the initial individual semantic networks—a complexity reduction feature we particularly value after having gone through the pain of interpreting the hairballs of full semantic networks comprising tens of thousands of nodes and links. Usually including 100-200 concepts, semantic networks of shared associations can be visually inspected as network diagrams and their use in the texts can be checked manually, which fits our mixed method approach.

## Meaning structures of two artistic collectives

Our analysis proceeds as follows. We begin by focusing on the overlap in the meaning structures of the two collectives. We expect meaning structures of artists across the collectives to be an outcome of their common belonging to the field of contemporary art. We: (1) retrieve concept associations shared by the artists[9] in Z and in C, separately; (2) detect concept associations overlapping between the collectives, i.e., associations shared in both; (3) locate and examine focal concepts and concept associations among those shared by the two collectives; (4) check the quotes using the focal concepts and associations to see if Z and C artists systematically use them to address the pivotal issues of the field, such as what is good or valuable within the field, what art is and what it is not, what art and artists should do.

We proceed by examining the focal concepts and associations idiosyncratically shared by artists in each the two collectives—and presumably related to their group-specific idiocultures— against the context of their group-specific practices. Namely, we utilize ethnographic and textual data gathered during two years of our ethnographic studies to check how the idiosyncratic meaning structures are used in verbal expressions of the artists to articulate their group idiocultures throughout various collective activities.

Finally, we apply our mixed-method approach to meaning structures of two subgroups in one of the collectives, C,—namely, a subgroup of artists and a subgroup of academic philosophers— and investigate if group meaning structures may result from confrontation of different social fields in a collective.

---

[9]  As our focus is on the artistic field, this comparison does not include C Academics.

### *Common meaning structure*

Fig. 3 presents the result of the intersection procedure that yields concept associations shared between the artists of C and the Z artists, only among Z, and only among C artists. Concept associations shared both by the members of Z and by the members of C correspond to meaning structures shared across the collectives and, perhaps, imposed on them by the field of contemporary art in which they both operate (solid lines in Fig. 3). Concept associations shared by at least three members of one collective but less than two members of the other collective are supposed to be collective-specific associations (dot-dashed lines for Z and dashed lines for C in Fig. 3); we shall consider these idiosyncratic associations in the next subsection.

_________________________________

Figure 3 about here

_________________________________

Very few of the associations are shared by both of the collectives, yet these few are most central to the meaning structure in Fig. 3 and the sum of the frequencies of these common associations comprises more than one third of the sum of all associations' frequencies (compare line widths in Fig. 3). In addition, the most centrally positioned common associations are used times more often than any of the non-shared associations. This indicates a strong consensus among the collectives on the very basic concept associations common to them. Let us consider the meaning structure that includes these associations.

Most of the focal concepts shared across the collectives (summarized in Table 2) are clearly referring to Russian contemporary art field: *art*, *contemporary*, *work (of art)*, *artist*, *young*, and *Russian*. Expectably, we find *art* to be the dominant concept: The two collectives connected this concept to other shared concepts using shared associations as many as 108 times in total, which is about one third more than the total number of shared associations with the second central concept *contemporary* (72). *Art* also has more unique shared associations (11) than any other concept in the meaning structure (the second being *artist*). The richness of associations with *art* corresponds to its pivotal role in the common meaning structure of the art field, being an abstract and flexible 'elaborated concept' with plenty of alternative meanings (see Yeung, 2005, p. 398) and thus capable of connecting different positions in the field.

_______________________________

Table 2 about here

_______________________________

Concept associations common to C and Z artists make sense upon minimal acquaintance with the (Russian) contemporary art scene. They depict *art* as *contemporary* and *political*, categorize *artists* as *good*, *young*, and *Russian*, point at the main museum of Russian art, and refer to *art—work*s (Fig. 3). The central association *contemporary—art* (Table 2) was used 72 times, which is almost 2.5 times more often than the second frequently used concept association *art—work* (29 times) and twice as much as the sum of all frequencies of all other common associations together. Clearly, *contemporary—art* is central to the meaning structure shared by the two collectives.

If Bourdieu is right in asserting that the core struggle within a field is the definition of what is good or valuable within the field, this focal concept association *contemporary—art* is to be systematically used in normative statements that express beliefs on what art is and what it is not, as well as prescriptions of what art and artists should do. To test this, we extract and inspect quotes containing all of the 72 (see Table 2) instances when the association

*contemporary—art* is used by the collectives' members.[10]  Indeed, we find both of the collectives to systematically make normative statements about *contemporary—art* in dozens of their expressions. Here is a couple of concise examples by Z, from among the many[11]:

> If one makes some things for sale, one can do this to earn, but I don't see here any **contemporary art**, which is meant to unveil problems. (ZD,[12] interview)

> I think it is like a conscious action. You may do something in **contemporary art**, but not posit yourself as a contemporary artist. It is this awareness that makes the difference. (ZO, interview)

C also utilizes *contemporary—art* to make normative statements. The most interesting example is, perhaps, the quote below. It captures a C artist, CB, articulating this perspective orally during an artistic intervention, followed by an approval by the rest of the collective, and another C artist, CG, narrates this in a written text:

> An abandoned corpus of a factory, resembling Stalker by Tarkovsky. Ruins of the industrial Soviet past. On the top, through the holes in the roof, the light is pouring in. On the concrete slabs, in the floods of the rainwater, moss and fern are greening; if you squat, there is an illusion that you are looking at a small-scale earth, again, as in Tarkovsky movies. Silent, like in a church. This is the place where one should do exhibitions of **contemporary art**, says [CB]. Everybody agrees. (CG, written text)

Such normative statements utilizing the common concept association *contemporary—art* systematically occur across written texts, interviews, and group conversations by Z and C artists. The written texts and conversations were not solicited by the researchers. The interviews were less unobtrusive, however, members of the collectives were willing and seemingly at ease making normative statements about art and artistic practices in the interviews, too.

Furthermore, expressions by Z and C artists using the association *contemporary—art* also appear to persistently reflect general beliefs common to the artistic field. The quotes below exemplify how this association is used to express one of these beliefs—that art is important to the spectator but it is not easy to understand:

> There should be something coming from people… Somehow, everyone should want. I mean, if a person has a wish to meet art, he will see it in the ***contemporary art*** too. The main thing in it is the eureka occurring. It is important for the viewer to have an opportunity to interpret. (ZF, interview)

> The thing is that nowadays, of course, visual art, ***contemporary art***, in particular, has predominantly become incomprehensible to the profane viewer, hasn't it? Ask yourself, what are these black hooks on the white [background], what are they telling you? And here, naturally, it is important to have an explanatory text by a critic or an artist, which explains his position. (ZC, interview)

> From this point of view it is necessary, of course, to make ***contemporary art*** a fashion. One should make more art projects involving people not from the art world. If I were the director of an institution, I would simply introduce a rule: do projects with people not involved in art at all. (CB, interview)

---

[10] The capability of our approach to extract only focal shared concept associations out of thousands of concepts and associations produced by the collectives comes in handy to gain the amounts of concepts, associations, and hence quotes manageable in qualitative analysis.

[11] Our qualitative analysis draws on inspecting hundreds of quotes (in Russian), where focal concepts and concept associations are put to use. We try providing as many examples of quotes as possible, but clearly, we cannot provide them all within a single paper. However, we are happy to translate quotes for a particular concept or concept association upon Reviewers' or Editors' request.

[12] Here and further: The first letter in the encoded member's name refers to the collective; the second letter was randomly assigned.

Inspection of the remaining quotes with the rest of the focal associations shared across the collectives (Table 2, 72 associations in total) yields similar results: The members systematically use these associations captured by our technique to reflect on what is good or valuable within the field. While we expected artists to generate idiocultures, our analysis so far seems to suggest it is more of a common stereotype, in line with the famous argument by (Bourdieu, 1996b): Taken together, art collectives tend to reproduce field-imposed meaning structure. Yet, is it so earth-shattering to find contemporary artists speaking about the norms and values of contemporary art? Our ambition in this paper is not to confirm the expectable, i.e., that both of the collectives converge on a common meaning structure that explicates field culture. The designation of our approach is in teasing out the less obvious.

And it does. Systematic comparison of the quotes reveals that although both of the collectives, indeed, use the common associations to make normative statements, in doing so each of them stresses its own specific perspective. For instance, consider once more the pivotal association *contemporary—art.* While both of the collectives use it to speak about what good art is and what artists should or should not do, C artists systematically instantiate this association to stress that the 'right' contemporary art is theirs, the critically engaged one, as opposed to the 'wrong' approaches, focused on the production of aesthetic objects (as well as all the rest, note the 'whatever'):

> [I]f you are not in this critical convention, then most likely you are making not contemporary things, not contemporary knowledge, not ***contemporary art***, but something else—cutting matryoshkas, knitting asses, whatever. Such an opposition: critical—affirmative. (CA, interview)

Another C artist uses the same association *contemporary—art* to justify the need for engaged art and his own career trajectory towards it that led him to joining C; on the way, the disengaged art and the gallery art scene are labeled as unproductive:

> Well, our project with him, more specifically, my part of it consisted in creating a graffiti poem that would take [CA]'s meanings as well as mine out of the exhibition space into the street space. At the time, I was doing <…> experiments with the public space, partially information practices, along with monumental art: graffiti, murals, various monumental things. At the moment, in 2003-2004, this was realized as a necessity. Genealogically, it was a natural move out of the local situation in Nizhniy Novgorod ***contemporary art***. These were all small and closed communities, unproductive in the sense that the meanings and knowledge were encapsulated in a very small circle, a very small social circuit. What lacked is taking all this into a space of presence, into society <…>. (CD, interview)

In contrast, for Z the aesthetics of objects produced is pivotal:

> I have works from the 90s, when I was drawing ravers with a blue pen, detailed ones, and when I show them to contemporary artists many also find these works interesting, so one can name them ***contemporary art*** too. <…> It is either a form, or a non-traditional approach to the material (ZO, interview)

Novelty is another issue pivotal to the perspective of Z on *contemporary—art*:

> I think it's a cool thing. Because there is a weight of these biennales, this distinction between young and not young. Forming the new ***contemporary art*** as a union of the new (ZF, conversation)

Most of the quotes with other focal common concept associations confirm this interpretation. For instance, consider *young—artist.* Involving many members who are in the early stages of their careers, Z tends to defend young artists:

> I think, it [Biennale of young art] is a cool thing <…> there should be limits for Kandinsky [a prize in Russian contemporary art given to young artists], one should not send [an application] after a certain age. ***Young artists*** there means under thirty five. (ZF, conversation)

> PRO ARTE [a foundation for contemporary art] is a kind of infrastructure. There, you can ask for money to do a project, they can give you a space. Not so long ago there was a discussion between artists and the curator of our group that **young artists** need studios and that PRO ARTE is thinking of organizing such a space. (ZO, interview)

In contrast, C, which consists of mature artists, persistently expresses skepticism towards the younger ones. Here is one of their leaders' quotes instantiating the same association:

> When I am teaching **young artists** I sometimes feel difficulties. Also, everything has become too easy: cinema, video, architecture in 3D, photography is transformed into sculpture. Everything looks great, however hollow, because there is no effort needed to manufacture. <...> I understand curators very well: So, here is a **young artist** who could be given a credit of trust. He participates one time, another, but there is no guarantee that after the third time in the project he won't leave the track. It is a tough question. (CA, interview)

To sum up, our ethnographic data reveals common meaning structures to be systematically instantiated differently across the collectives. Although at a glance the meaning structures shared by the two collectives seem to merely reproduce the field that functions as their primary frame of reference, qualitative analysis reveals that in practice meaning structures are instantiated to articulate specific perspectives of the collectives. Perhaps, Bourdieu (1993, p. 106) is right in saying that "distinctive signs produce existence in a world in which the only way to *be* is to be *different*, to 'make one's name', either personally or as a group"?

### *Collective-specific meaning structures*

The reader might have already noticed that the majority of concept associations in Fig. 3 are idiosyncratic to only one of the collectives (as many as 102 out of 118 unique associations are specific only either to C or to Z). Table 3 summarizes the most frequently used of these collective-specific associations and the concepts usually part of them.

_______________________________

Table 3 about here

_______________________________

Surprisingly, the majority of focal collective-specific concepts and associations are referring to the notions of the Russian art field, just like the common concepts and associations; consider: *art*, *gallery*, *biennale*, *artist, young, [a gallery in St. Petersburg][13], exhibition*, *[a gallery in St. Petersburg]—gallery, artist—union, young—biennale, gallery—space, creative—person* for Z and *art, Russian, artist, contemporary, space, public, public—space, art—history, creative—work, contemporary—artist,* for C. Little is idiosyncratic in these concepts and associations.

The only substantial vivid global difference between the meaning structures of Z and C artists is, perhaps, the cluster of associations around the concept *political* (Fig. 3). With no collective-specific associations of Z, this concept is among the focal for C (56, see Table 3), having more unique associations in their idiosyncratic meaning structure than any other concept but for *art* (Fig. 3). As the meaning structures plotted in Fig. 3 suggest, Z merely steps into the territory of political discourse together with C (solid line *political—art*), but then Z concept associations (dashed lines) are merely visible against the dense meaning structure C elaborated in the political domain (dot-dashed lines), including such associations as political—*movement, political—artist, political action, political—project, political history, work—movement, and collective—action* (Table 3). A collective's stance towards politics may be induced by the forces within the art field. Politically engaged art, such as pursued by C, claims less autonomy than the art by Z, which criticizes artistic compromise, would that be political or economic/market engagement. In field theory, the autonomous and heteronomous poles structure art fields (Bourdieu, 1983), so the collective-specific meaning structures concerning political engagement may be said to result from field forces. Especially in light of our observation that little is original in most of the focal concepts and associations of the collectives.

It is doubtful, however, that meaning structures only result from field forces that dictate meanings associated with a particular position in the field. As Bourdieu (1990) himself highlighted, practice is able to refract the impact of fields. To see this effect of practice, we argue, analysis is to be informed by in-depth ethnographic knowledge. Take another look at the meaning structure of Z (Fig. 3 and Table 3). Above, we concluded that most of its focal concepts and associations are common notions, clear to anyone who has at least some basic knowledge of the artistic field. However, in addition to these expectable focal meaning structures, some seemingly senseless ones are present. For instance, the concept *senior* is, surprisingly, as central to Z as *art*. Moreover, associations using this concept, namely *senior—art* and *senior—biennale* are the most frequently used ones in Z. These are easy to disregard as they look like data processing errors—but only unless one has spent months with Z, witnessing discussions of an ironic exhibition *'Biennale of Senior Art'* (mind that many Z members are over 50). Z invented this exhibition for a high-profile Moscow gallery in response to an invitation from a famous curator. The exhibition title mocks 'Moscow International Biennale for Young Art'[14]—a high-profile event that, just like the whole Russian contemporary art scene, prioritizes young artists as an asset that can be capitalized by curators in the art market. The discussions explicated the distinctive approach of Z, fundamentally based on irony towards the successful gallery art and mocking its authorities. Consider the following fragment of one of such discussions, very typical for the collective:

> ZC: 'Biennale of ***senior art'***. <…> There is irony about what contemporary art is. <…>
>
> ZC: By the way, it can be a special project! A parallel program! This is a good move!
>
> ZF: [laughing] There is a room, 22 meters: "Here is a parallel program! Its curator is"… <…>
>
> ZF: Sponsors, those who sponsor us. Labels.
>
> ZJ: Can we propose [a well-known Moscow curator's name] to be a curator of a single room? No more?
>
> ZF: Give a room to [another well-known Moscow curator's name]! [Laughing]. <…>
>
> ZF: And this goes to the smallest room. To the leftists. <…>
>
> ZJ: We are in the art field, sorry.
>
> ZF: Come on, we can quarrel. Nobody forbids quarrelling.
>
> ZJ: We are not spitting in someone's face, we are all in the field, doing art. It's a game.
>
> ZF: Absolutely. <…>
>
> ZJ: See, curator of one exhibition could be [a well-known Moscow curator's name], curator of the second one—[another well-known Moscow curator's name], and of the third one—Bedbug Vasiliy. So we find a bedbug and make it a curator—can't we? [ZF laughs].

---

[13] A highly visible gallery in St. Petersburg.

---

[14] http://youngart.ru/en/

To be sure, we also checked all of the remaining 20 instances when Z members use the concept *senior,* comprised of 11 instantiations of the focal association *senior—art* and of 9 instantiations of the second central association *senior—biennale.* All of the quotes discuss 'Biennale of Senior Art', fixing it as one of the central concepts for the emergent meaning structure of Z during the period. Note that the ironic biennale proposal was (perhaps, expectably) not accepted, so there is no way to properly interpret this association but for the ethnographic data, which is another argument in favor of the mixed-method approach.

If the correct interpretation of these focal concept associations is not possible without the ethnographic data, what about the other focal concepts and associations we so easily concluded to merely reproduce the artistic field? If we sensitize ourselves by the knowledge of the practical context of Z when considering the collective-specific associations between these focal concepts, they do not appear as trivial as they seem. *Art* for Z is *senior—art,* linked to the ironic *senior—biennale*. *Gallery* is associated mostly with *Z* to signify their own gallery space and with the broader gallery space, where the Z gallery is located. Initially, Z occupied a tiny corridor in the gallery, without a permission from the owners, and started exhibiting there. Such invasions and occupation of the bodies of cultural institutions are a distinctive practice which Z introduced to the Russian artistic scene; it is for these practices Z is mainly known. *Exhibition*, in turn, is exclusively *collective*, referring to a unique practice of collective exhibitions (see Fig. 4) that Z conducts every two weeks in the spaces they invade either continuously, like the gallery corridor or spaces in other cultural institutions of St. Petersburg, or on a one-time basis, such as a random grocery store or a public toilet. At these frequent collective exhibitions around city spaces, members are usually expected to present their recent works. As it is very intense for an artistic production process, lives of the members are revolving around the preparation of artwork(s) for the next instance of collective exhibition, hence the corresponding association is shared and persistent, used to discuss the original practice of the group. These interpretations are confirmed by inspecting each of the 40 instantiations of these associations across the collective practice of Z (as opposed to only 20 instances of the remaining focal associations of Z).

———————————————

Figure 4 about here

———————————————

This brief overview of Z focal concepts and associations in the context of our ethnographic knowledge suggests that most of the collective-specific meaning structures, even if they look trivial on the surface, express unique group practice and can be properly understood only in the context of this practice. The meaning structure of C is even less ambiguous in this respect: Most of the associations are clearly referring to the engaged leftist political art practiced by the collective: *collective—action, political—action, Soviet—union, and art—must*, *engaged—art, political—project, political—movement, work—movement, engaged—artist, Soviet—past, Soviet—time,* and *political—artist* (Table 3). These associations are so explicitly present, we argue, because two fields collide in C.

### *Meaning structures of Artists and Academics in C*

We label the two types of members in the C collective Artists and Academics. Artists have received professional training in the arts and art is their main occupation. In C, they promote an activist, politically and socially engaged perspective on art. Academics are affiliated with academic institutions, writing books and papers in political and social philosophy. Their perspective on art is grounded in political philosophy and history in correspondence with the principles of critical post- and neo-Marxist theory, the feminist theory, the Eastern tradition of postcolonial critics, and reflections upon the history of Western civilization, symbolic power of the state, citizens' rights, and freedom. Thus, two fields—the academic and the artistic—meet in C. This makes the collective a particularly interesting case to investigate meaning structures occurring when intense joint practice and interaction bring together different fields.

These two groups are also formally separated (Girvan & Newman, 2002) in their interaction networks (Fig. 5). The six Artists interact very frequently with each other but they interact less with the Academics, who also interact relatively frequently with each other, even though their level of interaction is lower than that among the artists. The Artists and Academics, then, are social groups in terms of their interaction patterns.

_______________________________

Figure 5 about here

_______________________________

The interviews show that the subgroups are recognized by C members themselves and there is a constant tension between the two. Artists criticize Academics and vice versa. Their criticisms concern the application of norms and values that are common to the other member's main field but foreign to their own field. The Academics criticize the Artists' use of theory and are skeptical about their emotional rather than rational behavior, thus reproducing dispositions and values associated with the academic field:

> There were contradictions because there are artists and non-artists <...> Those artists are not thinkers. They are not used to thinking rationally <…> we had a conflict because he started interrupting me and shouting when I was chairing a conference. Well, the way it is common to artists, but not in academia. (CC, Academic, interview).

The Artists stress the Academics' secondary role in artistic practice within the collective:

> Before we are on stage, everything is criticized, blamed, questioned, kicked. When we are on stage, I am the director, and there's my full totalitarian power <...> Because I have the vision I normally take it [power]. (CI, Artist, interview)

> <...> all the life of artists revolves around this production process, and it is more natural for us that we devote more time to collective projects than they [Academics] do. (CD, Artist, interview)

Are group structure and variation in members' primary field of occupation relevant to the meaning structure within the C collective? Are meaning structures affected by the confrontation of members of two different fields of symbolic production? To infer this, we mapped the structure of concept associations shared by at least two Academics or by at least three Artists in C[15] (Fig. 6).

\_\_\_\_\_\_\_\_\_\_\_\_\_\_\_\_\_\_\_\_\_\_\_\_\_\_\_\_\_\_\_\_

Figure 6 about here

\_\_\_\_\_\_\_\_\_\_\_\_\_\_\_\_\_\_\_\_\_\_\_\_\_\_\_\_\_\_\_\_

---

[15] We set a slightly higher threshold for Artists because there are more of them in C and hence the odds for their concept associations to overlap are greater. We also filtered out only the stable concept associations, i.e., those used more than twice.

Expectably, most of the associations are idiosyncratic to only one of the C subgroups representing the different fields. Meaning structures of C Artists (dashed lines in Fig. 6), similarly to Z artists, include common notions of the artistic field that revolve around the concept *art.* This concept has more unique associations than any other concept in the Artists' meaning structure (see Fig. 6) and is used times more frequently in the concept associations of Artists than any other concept (Table 4). C Artists are concerned with *art—history, good—artists, art— boundaries*, *pure—art, art—form.* Artists' meaning structure, thus, appears to reproduce dominant meaning structures of their field.

_______________________________

Table 4 about here

_______________________________

However, inspecting the quotes which put these focal concepts and associations to use we find Artists systematically (i.e., dozens of times by different persons) questioning artistic autonomy. The most vivid are the ones using concept associations that seemingly celebrate pure art to highlight the relativity of artistic boundaries instead of distinguishing pure art from engaged art, as the struggle for artistic autonomy would imply. Minding space limitations, out of a handful of quotes we selected two by different Artists to illustrate how this is usually done:

> Well, ***art boundaries*** have always been difficult to capture and were always constructed apophatically: "now art is not this and not that". Now, it is the same. I, for instance, smell it: this is art and that is not. On the other hand, change of the system, in twenty years: what we knew of art in the beginning of the 90s and what we know of it now are incompatible things. Five years pass and we'll see what the Arabs and the Chinese show. Now, an Arab foundation is created, it does an exhibition by Damien Hirst, it is business, money out of nothing: he can make as many of these skulls as he wants. See, who of the Chinese artists are sold well? Those are calligraphists, not Ai Weiwei. It's just total. Clearly, it was like that before, but I think, before, the understanding, the consensus on ***art boundaries*** was greater than now. All these applied things: design, social design,—should these be considered art or not? (CA, Artist, interview)

Compare the point on the ongoing global tendency towards artistic heteronomy that the Artist above puts forward by instantiating the focal association *art—boundaries* to the statement of another C Artist below. She draws on the local historical context of the post-Soviet transformations and her personal career trajectory to criticize the exaggerated idea of artistic autonomy still present in the Russian art field by utilizing the association *pure—art*:

> It was because of the 'perestroyka', and all of the artistic community was carrying the stigma of apoliticism, which was associated with the notion of pureness. Meaning, we, artists—pure ['pure' is in English in the original quote]. All the political is dirt. Dirt, dirt, dirt! And, in principle, this remains. <...> It was, of course, because of the Iron Curtain and that artists, and all the thinking people in general, and cultural.. Culture wanted to preserve itself and distance from the global political project 'USSR'. That is why a concept emerged: let us imagine there is *pure art*, let us imagine there is autonomy, that we concentrate our world ourselves and live in it and do it all. And that's how everybody lived. (CB, Artist, interview)

C Artists use the concept associations that seemingly refer to pure art to question it and its boundaries, which they present as imaginary, a phantom resulting from a historically developing defensive reaction of the art field on the Soviet pressure.

On top of that, Artists' meaning structure does contain some concepts and associations atypical for the Russian artistic mainstream. As we already know, the concept *political* is highly central for C Artists (see Fig. 6 and Table 4 and remember the peripheral position of this concept in the meaning structure of Z, see Fig. 3 and Table 3). In fact, it is one of their most central concepts, used more often than *artist*—one of the concepts most persistently imposed by the artistic field, as we have learned from Fig. 3 and Table 3. Moreover, among C Artists' concept associations is a number of leftist politically-flavored ones, such as *political—movement, engaged—art, Soviet— union, Soviet—past,* and *work*(er)*—movement*. It is clear to us that for the C Artists creative practice is inseparable from political practice, while at the same time their meaning structure often flags loyalty to the artistic field, firmly positioning themselves in it.

Academics' meaning structures (dot-dashed lines in Fig. 6 and Table 4) have little to do with the artistic discourse and signify belonging to the academic field. Unlike Artists, they do not share any stable associations with *art*. While C Artists, just like Z artists (remember the quote by ZF above), are concerned with *new—meaning,* for Academics *political—meaning* is of utmost importance. Their most frequent idiosyncratic concept associations draw on the notions of political theory (*private—life, free—time, place*(of)*—power, liberal—freedom, political—* *meaning,* and *leftist—tradition*) and refer to socio-historical context (*historical—event,* *historical—form,* and *intellectual—environment)*. These focal associations are supplemented with many others, contextualizing them in the academic discourse, such as *knowledge* (of)— *reality, revolution—crisis, academic—freedom, academic—environment,* and leftist theory discourse: *capitalist—reality, capitalist—system, work—place, bourgeois—society*, and so on.

Dozens of quotes by Academics confirm our interpretation. Here is a vivid example with one of the focal Academics' idiosyncratic concept links used in political theorizing; note that it is used to challenge artistic autonomy:

> Realizing one of its **liberal freedom**s—self-expression—in the space of the aesthetical, the artistic community lives with an illusion of an internal protest and revolution, objectively continuing to serve the capital and enrich it with new images. <…> The freedom of artist as a subject of actual history may be expressed not in a life of euphoric community in itself, but in action freeing the limits of the aesthetical, breaking the boundaries of community autonomy, directed to society in general, through political action. <…> Establishment of artistic autonomy too often looks only like a peaceful demonstration, controlled and restrained with a living wall of water jets and police batons. Even if art in its autonomy claims the right to make a distinction in society as a whole, it holds and fixes upon a position strictly allocated by power surrounding it. But what about violence which the peaceful demonstration is fraught with? In other words, can we expect that art breaks the conventions of contemporary society, regaining its lost significance, relevance? Following Foucault, we can show that autonomy is a natural result and the aim of regulatory disciplinary practices. (CE, Academic, written text)

Even when we find some atypical associations among the Academics' focal ones, such as *artistic—project* and *creative—action*, inspecting the quotes we find these associations used mostly to distinguish themselves from the Artists:

> Not being an artist, and not capable of being one no matter how much I would desire, because I do not think with artistic images, I cannot always join **artistic projects**. I am, for instance, more interested to participate in projects related to education or what you term "knowledge production"—seminars, conferences, summer schools" (C, Academic, interview)

Using the association *artistic—project,* one of the Academics even complains about a growing split between Artists and Academics, in which Artists are to be blamed, in his opinion. One might easily mistake this concept association for symbolizing integration between the subgroups, if not inferring the ethnographic context:

> <…> rather ***artistic projects***, they do not need our help now, they got the ball rolling… I still collaborate with the group because I find importance in the archive produced and the name we created together. In this sense, yes. Although there is not so much joint activity. (CC, Academic, interview)

Note that while Academics frequently reject being capable of doing art, we never witness the Artists saying they are incapable of reflecting on politics. On the contrary, they write political texts, as well as actively participate in and organize academic events. And they never blame Academics for doing political philosophy. This suggests Artists' meaning structures, exhibiting a significant degree of conformity to the artistic field on the surface, but in practice highlighting the engaged art perspective of C, may owe to the influence of Academics, whose meaning structures are affected by the academic field. These observations, however, do not tell us how the specific meaning structures of C collective emerged in practice.

Once again, knowledge of the context is crucial for understanding. Namely, knowledge of the historical development of the Russian artistic field and of the personal trajectories of C members within it. Fact is: When C members started their careers, politically engaged art in the present meaning of the term (i.e., as expressed by C concept associations and quotes) hardly existed in the field. Take another look at the quote by CB Artist above. All art in the USSR art scene was either engaged with the ruling communist party or autonomous. Politically engaged *critical* art could not be visible, only artistic propaganda. When the USSR collapsed, pure politically disengaged art finally became legal, while political engagement turned out to be stigmatized as such, especially the leftist engagement. C members were among the few who started advocating political engagement in the post-soviet Russian art and managed to establish it as a tradition in the Russian art field. It is something C Artists and Academics agree on:

> Since 2002, we have been saying that art is connected to politics, that contemporary art cannot exist without it. The word 'politics' back then, ten years ago, was something indecent, speaking of this was forbidden, inappropriate. Very few were saying "here are leftist ideas, Marxism", and this should be thought of <…>. From the beginning, we were in such an avant-garde movement, where we did not have publics, and we molded them ourselves. We appealed to an inexistent community (CA, Artist, interview).

> Now it is, especially after these meetings, a norm, but back then we were very lonely, isolated, everyone laughed at us. Or hated us, considering we want back to the USSR, or saying "you, the leftist, are only mimicking the West, we have no ground for that [politically engaged art] in Russia"—that's what all our friends said. (CC, Academic, interview)

C formed in 2003. It was one of the few leftist initiatives. It was happening when the Marxist or the leftist turn only started [in Russia]. Not even started, it was the beginning of the beginning, it was still a bit uncomfortable to speak some... to identify oneself with a leftist political project publically, even inappropriate. Everyone perceived this with bias and snobbism, coming from the nineties, <…> it was hard to find an environment to communicate in. <…> one of its [C] main functions was to create a disturbance in the Russian intellectual environment <…> There was such a liberal consensus in this art and intellectual environment. And legitimation of Marxist and generally anticapitalist expressions was an exceptionally hard task. It was work. We had to do a serious work to make people think that treating it [engaged leftist agenda] like that is a shame. Create some... to shake the bases once seemed firm. <…> If the ideas themselves become mainstream, it is very good, because it shows your efforts deliver result. <…> Such a cultural hegemony— an important part of our project, we want more of such groups, more discussions on these issues <…>. Our task is to engage as many people as possible, including the young ones. (C, Academic, interview)

Indeed, over the years, the collective, almost from scratch, created the engaged leftist art agenda, built an audience for it, and created an extensive community that includes dozens of Russian leftist artists, philosophers, art-critics, writers, social scientists and activists, a community that supports the core ideas and projects of C. In 2013, C also started a school of participatory art, the only of its kind in Russia, with dozens of artists graduating every year.

Then, no one was doing critical theory in Russia. Overall, the leftist ideology was a taboo. People were foe and misgiving towards it. <…> up in arms over our activity because we immediately announced our belonging to the leftist tradition and engaged artists' tradition. <…> St. Petersburg treated this as something profane: "politics—a dirty occupation". Some intimate, album-circle practices... There is an own truth in it, but we felt that when Putin came, a new epoch starts and roughly speaking, if you don't do politics, the politics will do you. <…> And for the ten years of our existence we promoted creation of a certain environment. <…> In St. Petersburg, there emerged an environment, a younger generation, more politicized and more like the western youth that is not scared of politics. Even if you do science, art, you do it consciously and understand how it influences the political situation in the county, the world. We did not have this for long. (CG, Artist, interview)

The meaning structure of engaged art emerged in the Russian art field to a large extent thanks to C. It is very much owing to C that we now see the concept *political* and the association

*political—art* among the focal in the field-imposed meaning structure (Fig. 3). This role of C is internationally acknowledged by art specialists (see, e.g., Roberts, 2010; Volkova, 2015), but perhaps most revealing is the fact that when Z members use the association *political—art* they tend to refer specifically to C:

> <…> in C they have this strict thing, sort of, well, not strict, but they have a philosophy, ***political art***. As far as I understand, it is an idea-based association. (ZD, interview)

Members of C point that it was interaction between Artists and Academics that enabled elaboration of the new meaning structure which C introduced to the artistic field. In the quote below, an Academic uses one of the shared concept associations that directly refers to *engaged—art*, in a reflection on how the meaning structure around art, politics, and knowledge emerged during a period of joint collective practices and interactions within C:

> On the one hand, we are between Piter and Moscow, on the other—between art and philosophy, theory. <…> first time the discussions were incredible, they were very hot. There were many themes related to discussion of what politically ***engaged art*** is, what real art is, what contemporary art is and how politics and art are related, as well as art and knowledge, and left and right, and so on. Everything was revolving around this triangle. Different people were the opponents, this circle kept changing, because one is taking this side, the other one—that side. These discussions are good because in the process a person often changes opinion. He argues, argues, defends something, and then comes to something else. So, somehow we passed this period and the project was established. (C, Academic, interview)

Artists admire these confrontations typical for C (as we know from CCs quote above, however, their style is not always appreciated by Academics):

> We have a conflict style of interaction, but I like it a lot. Because when you work in a compromise regime it is hard to be responsible for the result. And we decided in the C collective that for us result is a principal matter, and turned out to be pleased with it. Many of us for a long time were in a situation when the result is not so important, the process is crucial and bla-bla-bla.. But in this case you cannot realize yourself as an artist at all, completely. (CB, Artist, interview)

> During the first years of our existence, in the first century, we tried all means to overcome the division of labor based on the professional zones of responsibility. It was a unique and astonishing experience that we continue to embody, but now we are in a less intensive phase and.. every day we insist on this: we allow artists to be artists and philosophers—to be philosophers, because philosophy is better done by those who not merely philosophize, but does it at another level of understanding of the subject. And each of us, naturally,—and it is important—actively uses such verbal and writing practices. Every group participant at some moments writes and publishes some theoretical texts. (CD, Artist, interview)

Not an easy task (if possible at all), a complete convergence between the perspectives of Artists and Academics was never achieved, but the continuous search for convergence enabled cooperation, which even the skeptical CC admits:

Now, we have very different perspectives, and we never brought them together. But something in common was always found, some common orientation at something I roughly name avant-garde. <...> I used to write lots of dialogues in the beginning, because I wanted to show that there are different positions. Well, it is hard to find a common language with our artists, especially [CB], because the person thinks differently, does not have the corresponding education. Well, nevertheless, some cooperation was possible. (CC, Academic, interview)

The differences between arts and academy—and the corresponding conflict between Artists and Academics inherent in C—worked as the sources of creativity and novelty:

Well, they [Artists] influenced, mainly [CA] and [CD], because of their organizing activity [smiles]. They stimulated me to think of art, write about art. Maybe, I would not do it that much. For instance, we made an interesting [newspaper] issue about Brecht. For that I mastered Brecht's theory, otherwise I would not do that. These things take place. It was also interesting to take part in artistic actions <...> naturally, it affects me because some creative potencies develop. (CC, Academic, interview)

One of the Academics depicts how this creative collision between Artists and Academics, purposefully overcoming the boundaries between art and philosophy through interaction within the collective, enabled the collective to produce its meaning structure, introduce it to the field, and maintain it in a hostile environment; very much in line with how Farrell (2003) describes creative circles enable producing and promoting new artistic ideas:

All this directionality of the project; it is related to overcoming of the artistic or collective's locality <...> between art and philosophy, theory <...>. There were lots of topics involving discussions of what art is, what is real art, what the contemporary art overall is, how arts and politics are related, art and knowledge, right and left, and so on. And everything revolved around this political triangle. Legitimation was, indeed, not easy, there were many obstacles on the way and we had to struggle for it, but the struggle was conscious. That's the sense of the collective—it has a certain solidarity, and with a feeling of one's political and aesthetical righteousness this collective mechanism works. <…> For instance, we have a project that exists for several years. An attempt, in a short period of time, say two days, to create such a space where the boundaries are overcome of, first, own ego, individuality, through the collective; Boundaries between activism, theory, and art are overcome through interaction with each other. <…> I remember [CC, Academic], was writing some plays and together we invented some actable dialogues. I tried something similar, some poetic genre <…> and it was interesting to propose this to other authors. You tell them: write for the [C] newspaper. Some philosopher. <…> same for artists and theorists. Plus, there may occur more confronton, because sometimes something may be unclear to us or to them, because the method is different. Well, even the way to discuss is sometimes different. But sometimes something unexpected emerges from these discussions. Some funny product, or a play, some dialogue. Contradictions are not a destructive element, on the contrary, constructive. <…> the disciplinary boundaries, I still think, are a huge shame. Everyone must be a bit of an artist, try; everyone must try to be a little of a theorist. <…> I think this union only makes sense when this mutual interference makes both of the fields—art and science—better <…> It is not even a synthesis, but contradiction or, even more, a problematization, because we make all these conflicts explicit, sometimes absurd, and work them through aesthetically and intellectually. (C, Academic, interview)

Based on this qualitative analysis, we are convinced that the cooperation between Artists and Academics in C has played a decisive role in introducing the meaning structure related to political engagement to the Russian art field. More precisely, our previous analysis suggests that Artists in C embraced Academics, and with them—the idea of political engagement, heteronomously adjusting the culture of their own field accordingly. However, those are Artists, not Academics, who idiosyncratically share the association *political—art,* now common in the artistic field (Fig. 3). So perhaps, influenced by the Academics, those were the Artists who, in the end of the day, introduced the idea of engaged political art to Russian artistic field.

Resulting from the confrontation between the two fields in C is the meaning structure shared between Artists and Academics, represented by solid lines in Fig. 6 (the corresponding focal concepts and concept associations are summarized in Table 5). From the table and the figure it is clear that over the years of confrontation Artists and Academics achieved quite some common ground and a common perspective on what politically engaged leftist art is. This created a kind of interface between the two fields within the C meaning structure, visible in the plot as the vertical solid line between the meaning structures of Artists and of Academics and explicating the core ideas promoted by C: *leftist—movement*, *collective—action*, *civil—society, political—* *action, new—forms, contemporary—artists, subject—(of)history, political—situation historical—moment,* and *new—world.*

\_\_\_\_\_\_\_\_\_\_\_\_\_\_\_\_\_\_\_\_\_\_\_\_\_\_\_\_\_\_\_

Table 5 about here

\_\_\_\_\_\_\_\_\_\_\_\_\_\_\_\_\_\_\_\_\_\_\_\_\_\_\_\_\_\_\_

As we already know from the previous analysis, the essence of C idioculture, distinguishing the collective from the rest of the artistic field, is explicated by their focal concept *political.* Fig. 6 illustrates this concept is the pivotal point of the meaning structure mediating between the two fields that meet in C. This concept has more unique associations than any other concept. Roughly a third of these associations are shared by Academics, a third shared by Artists, and a third—by both Academics and Artists. *Political* is clearly the main point where meaning structures of the academic and artistic fields meet in C. This makes it an 'elaborated concept'. In the study by Yeung (2005, p. 398), such concepts connect urban communities. Our findings suggest they may also connect different social fields.

Four associations with Political are shared both by Artists and by Academics: *political—action*, *political—situation*, *political—project,* and *political—issue*. In the 24 instantiations of these concept associations we find the evidence of a stable substantial resonance between Artists and Academics of C in how they address contemporary art. This is the essence of their idioculture. And in the ways these associations are used we find a mutual movement of Artists and Academics towards understanding each other, perhaps a sign of mediation between the fields. Consider examples of how a C Artist instantiates the association *political—situation* to criticize the supposed autonomy of art:

> I think, it all started when the **political situation** in the country and the world began to change. When the ongoing events could not be ignored any longer. And then European projects started to grow stronger. This is also related to crisis; the whole world was in crisis. Altogether, more and more people came to understand that art can't move further while being positioned in an absolute sovereignty, in which it had been before. What is the issue now? The main debate is along the following line: May an artist insist on his sovereignty, or should ways be searched in

> alternative projects? And if you work alone, you are someone else, rather an agent (CB, Artist, interview).

An Academic, in turn, uses the same association *political—situation* to not only highlight the importance of engaged art, but also to acknowledge the limitations of academic philosophy that the engaged art is capable of overcoming. The form is also revealing: To express this point, she speaks the mouth of an artist as part of a fiction dialogue between an activist, an academic and an artist, written based on one of the intense discussions between Artists and Academics we described above:

> Activist: When I read biographies of Italian labor activists written 50 years ago, I am impressed with the co-incidence of the existential dimension of personal life and the large history. These two levels could not be split. People took personal decisions – moved from one city to another, married, gave birth to children, changed jobs and names – based on what was happening in the world, what was the ***political situation*** and what the collective struggle demanded from them. And where is this all now? Everyone is saving his own ass, and then theorists cry that in London another philosophical faculty has been closed and professors were fired because of the economic crisis and unprofitability.

> Artist: There is nothing to cry about. To me, philosophical faculties, if not closed, should have eliminated themselves, because academia is something obsolete, medieval in essence. All the system of academic hierarchy is organized or the person to be driven into the press of scholarly monotonousness and his career path is covered with book dust. Philosophers do not invent new forms, they do not give a damn about collectivity. (C, Academic, written text)

We conclude that the shared meaning structure of C rather emerged from the joint practice of Artists and Academics in the collective than was imposed by any of their fields. Such position-taking is, perhaps, what introduces changes to fields. The cooperation of Artists and Academics in C promoted a focus on political engagement in the Russian artistic field and gave rise to the original meaning structure criticizing artistic autonomy that changed the field. To locate such focal meaning structures, quantitative analysis is helpful. Understanding them requires qualitative analysis based on the ethnographic knowledge.

## Conclusion

Both field and institutional perspectives intertwine social and symbolic structures, mapping the space of behavior, properties, and ties on the space of meanings (Bourdieu & Johnson, 1993;

Mohr, 1994; Mohr & Duquenne, 1997; Mohr & Neely, 2009). We very much value the ambition. We also subscribe to the idea that fields and institutions impose systems of classification on their members that resist change and attempt to regulate the meanings that members attribute to behaviors, properties, and ties. However, these perspectives seem to underestimate the role of groups operating in social and institutional fields and putting the relationship between cultural meanings and social structure into practice. Drawing on the capability of groups to generate group-specific cultures (Fine, 1979; 1991; Yeung, 2005; Fine, 2012), we conceptualize the dual ordering of the social and the cultural not as the property of an individual person but as the property of a small group. We define a group as a set of persons involved in regular and engaging interactions. While interacting, members may be confronted with meanings assigned to behaviors, properties, or ties that differ from their own. If interaction is regular, engaging, and important to the members, it can be consequential to the meaning structure.

To sensitize analysis to meaning structures in the social practice of small groups we proposed a mixed-method inductive socio-semantic network approach. Namely, we started with computer-assisted processing of free narratives by individuals to map meaning structures as semantic associations shared in a group. Then, the focal elements were quantitatively detected in these group-specific meaning structures—to guide our attention to the cores of the group cultures while avoiding interpretation bias. This way focal group meaning structures are derived inductively, without selecting the 'keywords' researchers or extragroup experts consider to be important prior to counting, but at the same time allowing to find the central elements. This approach makes a contrast both to other inductive automated approaches to culture that do not yield context-specific associations between cultural elements, such as topic modeling (Blei et al, 2003; DiMaggio et al, 2013; Mohr & Bogdanov, 2013), and to approaches used by institutional and field theory scholars that formalize culture via category coding (Mohr, 1994; McLean, 1998; Mohr, 1998; Mohr & Lee, 2000; McLean, 2007; Mohr & Neely, 2009). However, we did not stop there and examined how the focal elements of the meaning structures were instantiated in expressions of the members, supplementing this analysis with other ethnographic data available to us.

Our results highlight the importance of mixing methods in socio-semantic network analysis. Quantitatively analyzing a combination of textual and social network data on two St. Petersburg art collectives, we initially found their meaning structures to reproduce the field, referring to its pivotal issues even though artists are generally expected to express original meanings. However, a qualitative examination making use of our extensive textual and ethnographic data revealed that collective meaning structures, whether they seem to be imposed by the field or not, are substantially impacted by collective-specific practice. Collective meaning structures either are instantiated differently in line with peculiarities of collective practice or even emerge from this practice. Thus, although finding central elements in group-specific semantic associations points us to the core of its meaning structure, it is not sufficient to properly understand (*Verstehen*) these meaning structures.

The main limitation of our approach is, perhaps, that the produced concept maps per se do not reveal group-specific meaning structures and hence cannot be the final stage of analysis. In contrast to the ones produced via coding or by formulating closed-ended survey questions, they do not seem as 'important', or 'interesting' at a glance. Even having done ethnographies in the two collectives under study for many months, we had to systematically work through loads of ethnographic notes and quotes before we could understand the relevance of a certain concept association to the meaning structure. Hence, computational mapping and quantitative analysis of the meaning structures cannot substitute the qualitative efforts, but only directs qualitative analysis that is to tease out the contextual meaning. If a researcher does not have access to such in-depth data, like in the case of historical data (Mohr, 1994; McLean, 2007), inductive mapping may be misleading—just as our mixed-method analysis shows.

Another limitation is that the maps produced using the proposed technique are not necessarily relevant at all times, even in the same local context. These maps are practice-driven, and practice is fluid (Bourdieu, 1990). For instance, during the period of data collection, the concept *senior* was so important for Z, but it does not mean the concept will be so central even in the very same collective in a year, when the corresponding 'Senior Biennale' passes. Most of the group-specific maps capture meaning structures that were emerging at the time of data collection—and should be treated as such.

Furthermore, to examine local group effects on meaning structures, we have purposefully selected a rather peculiar setting: artistic collectives. It has been argued that a group's coherent artistic style, techniques, and topics emerge thanks to discussions between members who form a shared 'vision' (Farrell, 2003). The selected collectives strongly encourage collaboration, so there is a lot of interaction going on and the interaction is quite engaging. Interaction, then, may affect meaning structures in artistic collectives and, perhaps, other collectives pursuing common goals, like start-ups, academic departments, research groups, activist groups, more strongly than in more formalized settings. This limits our findings to such kinds of groups. In addition, creative practice as studied here is not formally regulated, in contrast to bureaucratic practice, such as encountered in a welfare organization (Mohr, 1994) or a city administration (Meyer et al, 2012). Within highly regulated practices, it would be much more difficult to pinpoint local meaning structures.

Bearing in mind all the limitations involved, we are convinced that in order to study cultural practice a departure is necessary from the nicely polished cultural structures to the messy structures of locally emerging culture. We tried to show that it does not necessarily mean ending up with senseless 'hairballs' in one's hands. One way to reconcile between 'big' and 'thick' data approaches, which is being called upon (Bail, 2014; Breiger, 2015; Latzko-Toth et al, 2017), is through a mixed method-based *Verstehen*.

# Local Meaning Structures

**Tables and Figures**

Table 1. Overview of the full semantic networks.

| Member | Type | Concepts | | | | Concept associations | |
|--------|------|--------------|-----------|-----------------|--------|--------|--------|
| | | *Conversation* | *Interview* | *Written text* | *Total* | *Total* | *Shared* |
| CA | Artist | 0 | 3395 | 9270 | 12665 | 6918 | 839 |
| CB | Artist | 0 | 2901 | 0 | 2901 | 1255 | 197 |
| CC | Academic | 0 | 1646 | 20722 | 22368 | 12223 | 958 |
| CD | Artist | 0 | 1920 | 2614 | 4534 | 2405 | 359 |
| CE | Academic | 0 | 0 | 8220 | 8220 | 4826 | 499 |
| CF | Artist | 0 | 0 | 6668 | 6668 | 3994 | 434 |
| CG | Artist | 0 | 3965 | 27276 | 31241 | 17911 | 1171 |
| CH | Academic | 0 | 2453 | 2989 | 5442 | 2725 | 427 |
| CI | Artist | 0 | 928 | 0 | 928 | 356 | 79 |
| ZA | Artist | 0 | 536 | 0 | 536 | 165 | 40 |
| ZC | Artist | 1656 | 2055 | 0 | 3711 | 1296 | 216 |
| ZD | Artist | 0 | 1813 | 0 | 1813 | 563 | 113 |
| ZE | Artist | 0 | 1288 | 0 | 1288 | 437 | 79 |
| ZF | Artist | 1307 | 920 | 0 | 2227 | 770 | 119 |
| ZG | Artist | 264 | 0 | 0 | 264 | 106 | 16 |
| ZI | Artist | 369 | 3551 | 0 | 3920 | 1434 | 182 |
| ZJ | Artist | 744 | 2664 | 782 | 4190 | 1509 | 210 |
| ZL | Artist | 853 | 0 | 0 | 853 | 289 | 46 |
| ZN | Artist | 0 | 1933 | 0 | 1933 | 556 | 123 |
| ZO | Artist | 0 | 3355 | 0 | 3355 | 1193 | 198 |

*Note: The first letter in an acronym for a member refers to the collective. The numbers under Conversation, Interview, and Written text give the number of concepts collected for a member from texts of the specified type. Under Concept associations, Total shows the number of distinct concept pairs for this member and Shared Associations represents the number of these associations also used by at least one other member of the two collectives.*

Table 2. Concept associations, shared both in C and in Z, and concepts most frequently used in the shared associations.

| Concept | Co-occurrences with other concepts | Concept association | | Frequency of usage |
|---|---|---|---|---|
| art | 108 | contemporary | art | 72 |
| contemporary | 72 | art | work | 29 |
| work | 29 | young | artist | 13 |
| artist | 25 | good | artist | 7 |
| young | 18 | political | art | 7 |
| Russian | 11 | Russian | museum | 6 |
| good | 7 | Russian | artist | 5 |
| political | 7 | young | person | 5 |

Table 3. Concepts, most frequently used in the idiosyncratic associations of Z and C artists, and most frequently used idiosyncratic concept associations.

| Z | | C | |
|---|---|---|---|
| *Concept* | *Co-occurrences with other concepts* | *Concept* | *Co-occurrences with other concepts* |
| senior | 21 | art | 85 |
| art | 21 | political | 56 |
| gallery | 19 | history | 50 |
| biennale | 17 | action | 46 |
| artist | 9 | Russian | 46 |
| person | 9 | artist | 42 |
| [a gallery in St. Petersburg] | 9 | collective | 42 |
| young | 8 | contemporary | 40 |
| exhibition | 7 | space | 36 |
| space | 6 | public | 36 |
| *Concept association* | | *Frequency of usage* | *Concept association* | | *Frequency of usage* |
| senior | art | 12 | public | space | 18 |
| senior | biennale | 9 | art | history | 14 |
| [a gallery in St. Petersburg] | gallery | 9 | creative | work | 12 |
| young | biennale | 6 | collective | action | 11 |
| [Z] | gallery | 5 | Russian | language | 9 |
| collective | exhibition | 5 | contemporary | Russia | 9 |
| artist | union | 5 | political | action | 8 |
| good | text | 3 | contemporary | artist | 8 |
| gallery | space | 3 | art | must | 7 |
| creative | person | 3 | Soviet | union | 7 |

Table 4. Idiosyncratic concept associations, most frequently used by C Artists and Academics, and most central concepts in their idiosyncratic semantic networks.

| Artists | | Academics | |
|---|---|---|---|
| *Concept* | *Co-occurrences with other concepts* | *Concept* | *Co-occurrences with other concepts* |
| art | 33 | life | 14 |
| history | 17 | political | 13 |
| political | 14 | event | 13 |
| artist | 10 | historical | 12 |
| movement | 9 | environment | 12 |
| Russian | 9 | meaning | 10 |
| private | 8 | time | 9 |
| Soviet | 8 | reality | 9 |
| engaged | 7 | form | 8 |
| contemporary | 7 | place | 8 |

| *Concept association* | | *Frequency of usage* | *Concept association* | | *Frequency of usage* |
|---|---|---|---|---|---|
| art | history | 13 | historical | event | 7 |
| good | artist | | private | life | 6 |
| political | art | 5 | art | project | 6 |
| Soviet | union | 5 | free | time | 6 |
| political | movement | 5 | intellectual | environment | 6 |
| private | property | 5 | historical | form | 5 |
| Russian | situation | 5 | place | power | 5 |
| art | boundaries | 5 | liberal | freedom | 4 |
| art | form | 4 | political | meaning | 4 |
| engaged | art | 4 | leftist | tradition | 4 |

Table 5. Shared concept associations, most frequently used by C Artists and Academics, and most central concepts in their shared semantic network.

| Concept | Co-occurrences with other concepts | Concept association | | Frequency of usage |
|---|---|---|---|---|
| movement | 30 | leftist | movement | 21 |
| leftist | 26 | new | form | 17 |
| political | 24 | historical | moment | 12 |
| new | 24 | collective | action | 11 |
| form | 22 | subject | history | 9 |
| action | 19 | civil | society | 9 |
| collective | 16 | contemporary | artist | 8 |
| society | 15 | political | action | 8 |
| contemporary | 14 | political | situation | 7 |
| time | 13 | new | world | 7 |

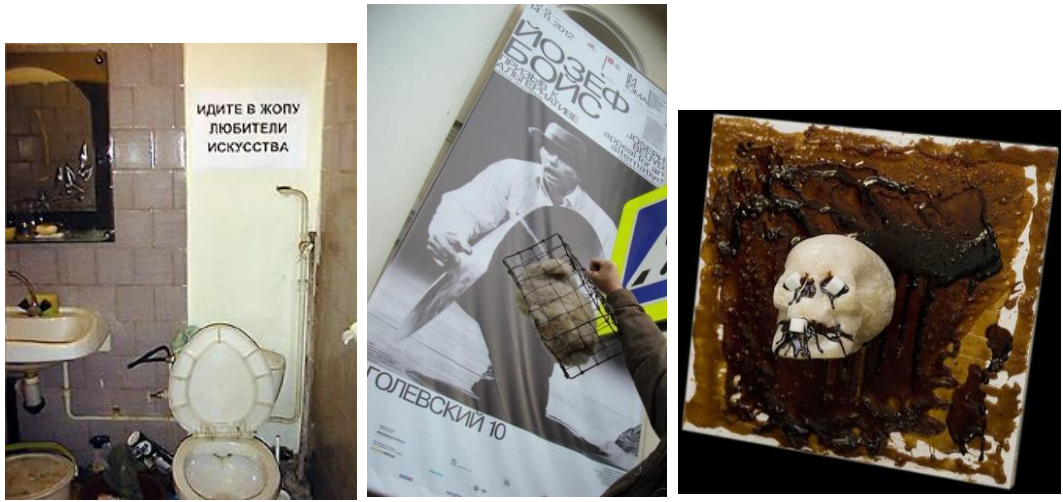

**Fig. 1** Exemplary artworks by Z

*Note: Left to right: a performance at J. Beuys' retrospective exhibition in Moscow in 2012, which consisted in visiting the exhibition with a fake white a-la-Beuys but caged hare, mocking the 1965 performance by Beuys 'How to Explain Pictures to a Dead Hare' at his first personal exhibition; 'Eighth Toilet Exhibition' in a toilet of an art center, consisting in the sign 'Go to ass, art lovers'; a parody on D. Hirst's skull, playfully attributed not to Z, but to Hirst himself.*

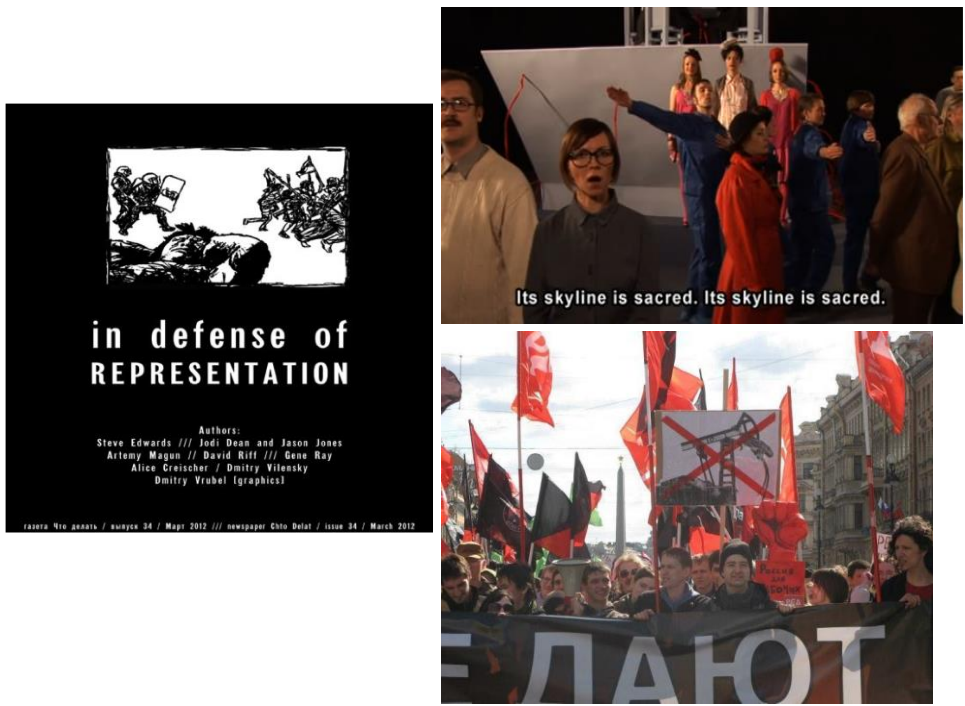

**Fig. 2** Exemplary artworks by C

*Note: From the left, clockwise: a cover of one of the newspaper issues; a frame from a Songspiel against the construction of Lakhta Centre (also known as the 'Gazprom tower') in St. Petersburg (the building would disrupt the skyline of the historic centre of St. Petersburg—a UNESCO World Heritage site, hence the construction plan evoked a huge protest movement, which C contributed to); performance 'Russian Forest' framed as a 1st of May demonstration on the Nevsky prospect, the main street of St. Petersburg.*

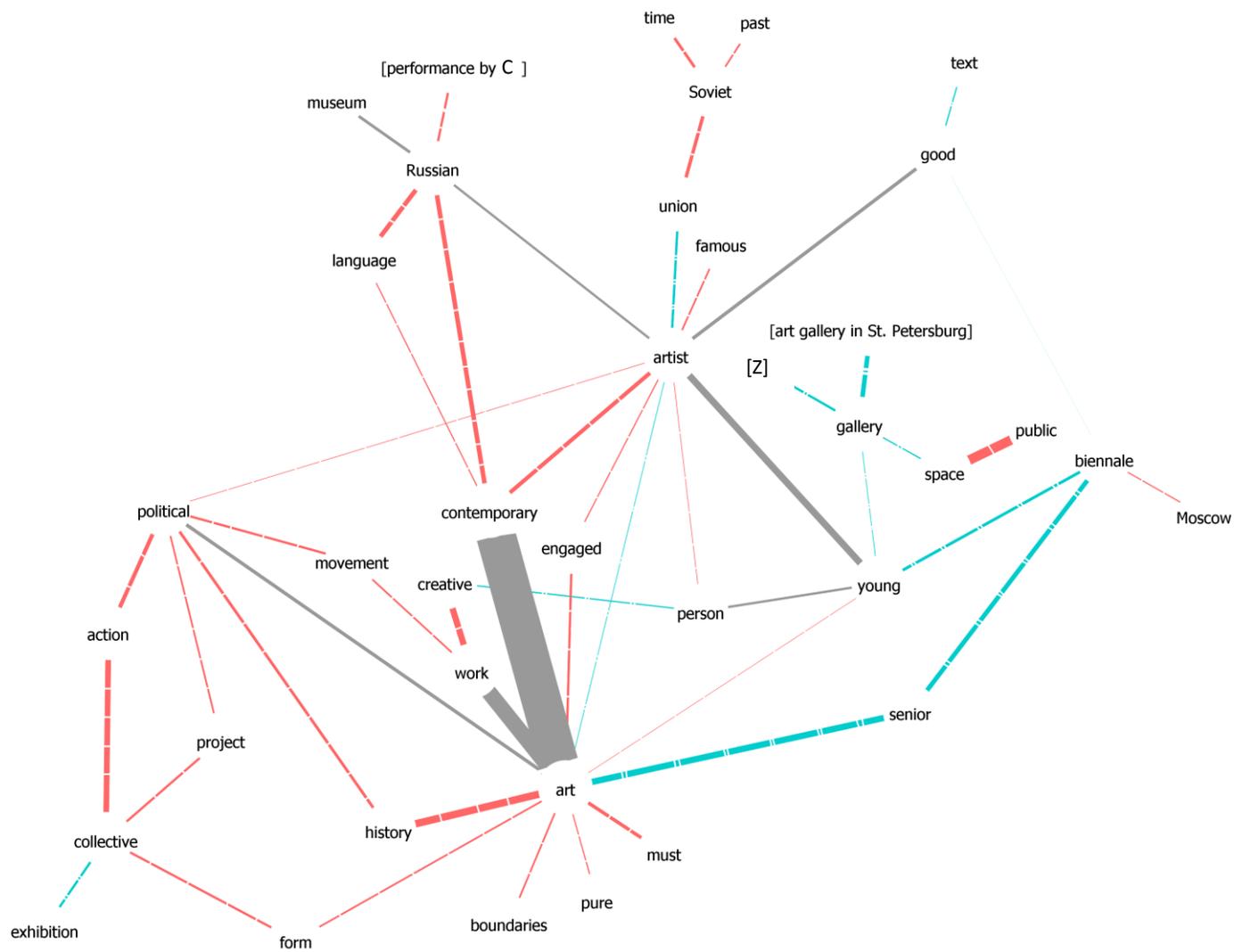

**Fig. 3** Principal component in the network of shared concept associations within and across artists of C and Z

*Note: Concept associations specific to Z: dot-dashed lines; concept associations specific for C: dashed lines; concept associations shared by both collectives: solid lines. Line width corresponds to the number of times a concept association was used.*

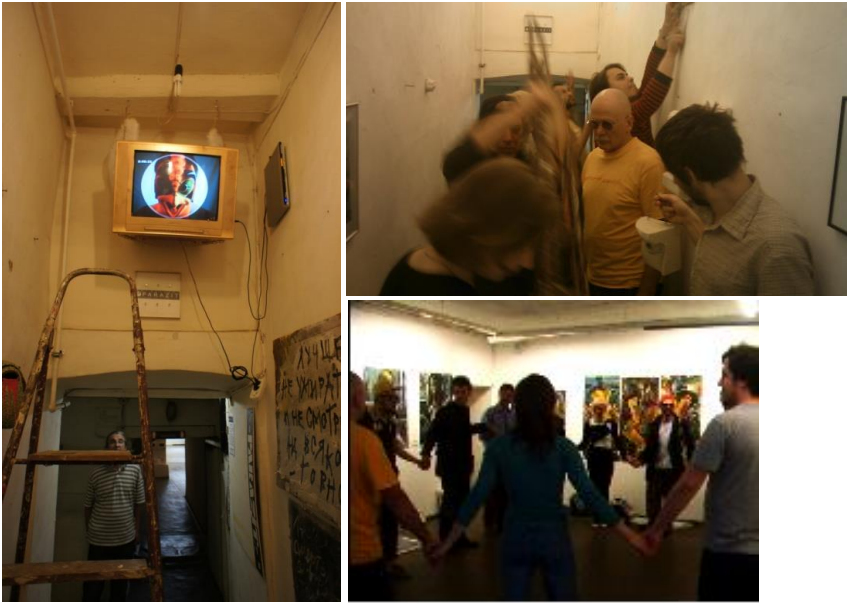

**Fig. 4** Z preparing its collective exhibitions

*Note: From the left, clockwise: corridor of the Z gallery before exhibition preparation; Z preparing an exhibition in the corridor; Z hand-holding ritual circle after an exhibition had been prepared.*

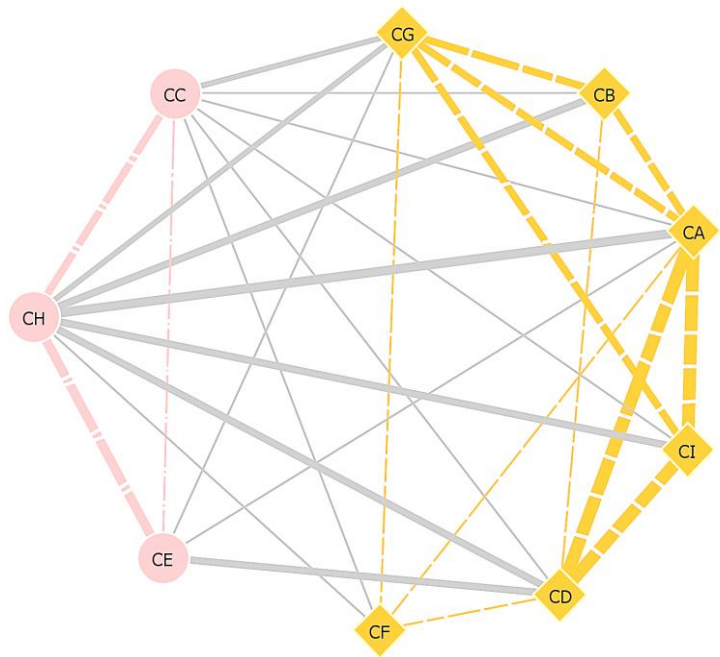

**Fig. 5** The structure of interpersonal relations (interaction) among C members

*Note: Circles: Academics. Squares: Artists. Dot-dashed lines: ties between Academics. Dashed lines: ties between Artists. Solid lines: ties between Artists and Academics. Line widths indicate quantified interaction frequency scaled from 'interacting more than once a month' to 'interacting every other day or more often'.*

**Fig. 6** Principal component in the semantic network of shared concept associations within and across Artists and Academics subgroups in C

*Note: Dot-dashed lines: concept associations of Academics; Dashed lines: concept associations of Artists; Solid lines: concept associations shared by Artists and Academics.*